\documentclass[12pt]{article}
\usepackage{graphicx}
\usepackage{subfigure}
\usepackage{amssymb}
\usepackage{amsfonts}
\usepackage{latexsym}

\usepackage[usenames]{color}
\usepackage{latexsym}
\usepackage{epstopdf} 
\usepackage[normalem]{ulem}

\newcommand{\beq}{\begin{equation}}
\newcommand{\eeq}{\end{equation}}
\newcommand{\be}{\begin{equation}}
\newcommand{\ee}{\end{equation}}
\newcommand{\bea}{\begin{eqnarray}}
\newcommand{\eea}{\end{eqnarray}}
\newcommand{\beas}{\begin{align*}}
\newcommand{\eeas}{\end{aligin*}}

\makeatletter

\@addtoreset{equation}{section}
\makeatother

\usepackage{amsmath}	
\begin{document}


\begin{titlepage}
\begin{flushleft}
       \hfill                       FIT HE - 23
-01 \\
       \hfill                       
\end{flushleft}

\begin{center}
  {\huge Instability of holographic cold compact star \\
   \vspace*{2mm}
with color superconducting core 
\vspace*{2mm}
}
\end{center}

\begin{center}

\vspace*{5mm}
{\large ${}^{\dagger}$Kazuo Ghoroku\footnote[1]{\tt gouroku@fit.ac.jp},
${}^{\dagger}$Kouji Kashiwa\footnote[2]{\tt kashiwa@fit.ac.jp},
Yoshimasa Nakano\footnote[3]{\tt ynakano@kyudai.jp},\\
${}^{\S}$Motoi Tachibana\footnote[4]{\tt motoi@cc.saga-u.ac.jp}
and ${}^{\ddagger}$Fumihiko Toyoda\footnote[5]{\tt f1toyoda@jcom.home.ne.jp}
}\\

\vspace*{2mm}
{${}^{\dagger}$Fukuoka Institute of Technology, Wajiro, 
Fukuoka 811-0295, Japan\\}
\vspace*{2mm}
{${}^{\S}$Department of Physics, Saga University, Saga 840-8502, Japan\\}
\vspace*{2mm}
{${}^{\ddagger}$Faculty of Humanity-Oriented Science and
Engineering, Kinki University,\\ Iizuka 820-8555, Japan}
\vspace*{3mm}
\end{center}

\begin{center}
{\large Abstract}
\end{center}

We study a holographic model of quantum chromodynamics, 
which can describe a color superconductor and a dilute nucleon gas phase. 
The two phases are adjoined in the phase diagram at a critical value of the chemical potential.
In other words, a first-order transition from the ordinary nucleon gas to the color superconductor (CSC) is found by increasing the chemical potential.
This model is suitable to investigate the possibility of a cold compact star with a color superconducting core. 
The equation of state of the star is given by the holographic model considered in this article, and we find that it is 
impossible in the present model 
to find a hybrid star of nuclear matter and the CSC core through the relation of mass and radius of the star by solving the Tolman-Oppenheimer-Volkoff equations.
Several other interesting implications are given by using the EoS.

\noindent

\begin{flushleft}

\end{flushleft}
\end{titlepage}
\newpage

\section{Introduction}

The gravity/gauge duality \cite{Maldacena:1997re,maldacena1999large,Gubser:1998bc,Witten:1998qj} is very useful to investigate the quantum chromodynamics (QCD) in its strong coupling regime.
In particular, this approach is powerful to investigate the thermodynamic system with the baryon number chemical potential ($\mu$) since this approach is free from the sign problem. In this approach, the Gibbs free energy has been examined, and then the phase diagram can be found in the $\mu$-$T$ plane. 
We expect to find color superconductivity (CSC) in QCD at large $\mu$.

At first, a holographic way to investigate superconducting condensed matter has been opened by the authors of \cite{Gubser:2008px, Hartnoll:2008vx}, where a bottom-up model has been proposed.
This model has been extended to the theory with superconductivity of other kinds of charges \cite{Nishioka:2009zj, BitaghsirFadafan:2018iqr, Basu:2011yg}, and also to the CSC in QCD \cite{Horowitz:2010jq, Evans:2011eu, Ghoroku:2019trx}.

In the case of QCD, within the probe approximation for the flavor matter system, 
we can find the phase diagram which provides five regions in the space of the temperature and the chemical potential.%
\footnote{Here we concentrate on the situation where the flavor part can be regarded as a probe under some appropriate conditions, for example, $N_f\ll N_c$.
}
Two of them correspond to the CSC phases, one is in the high temperature deconfinement phase \cite{BitaghsirFadafan:2018iqr} and the other is in the low temperature confinement phase \cite{Nishioka:2009zj,Horowitz:2010jq,Ghoroku:2019trx}. 
Therefore, the CSC phase is realized in both the confinement and deconfinement phases for sufficiently large $\mu$. 

{We notice that our bottom up model is suggested by the Type IIA superstring theory at
large $N_c$, the number of the stacked D$_4$-branes, in which the Yang-Mills gauge fields live. 
As for the flavored quarks, which are introduced as $N_f$ D$_8$-branes, are supposed as the probe for $N_f\ll N_c$. Then
they are neglected in setting the ground state of QCD. In other words, the backreaction from the dynamical quarks
are neglected to find the ground state of the theory. 
In our model, the probe field, which is dual to color non-singlet scalar operator of QCD, is set in the bulk.
This point should be noticed since such bulk fields dual to color non-singlet operators are not introduced usually.
We dared to adopt such a model to investigate CSC phase further.
Within this probe approximation, on the other hand, we could find 
the CSC phase at large chemical potential even in the confinement phase.}

In this modeling, at low temperature side, the CSC phase is found for $\mu>\mu_\mathrm{t}$ \cite{Nishioka:2009zj, Horowitz:2010jq}. 
Here, we notice that the region $\mu<\mu_\mathrm{t}$ is furthermore separated into two, one corresponds to the baryon matter phase for $\mu_c<\mu<\mu_\mathrm{t}$ \cite{Ghoroku:2012am,Ghoroku:2021fos} 
\footnote{{In Ref.\,\cite{Ghoroku:2012am}, $D_4/D_8$ top-down model is used, but we here use the instanton model given in Ref. \cite{Ghoroku:2021fos},
where a bottom-up model is used in the same footing with the one of the CSC model.}}
and the other to vacuum phase for $\mu<\mu_c$.
The vacuum can be considered as an insulator with respect to the baryon charge current since there is no charge density to form the normal current. We notice that, in the case of the baryonic matter phase, the baryonic charge density exists, however, there is no color charge density, which is generated through a phase transition in low temperature confinement phase.

How can we see such a CSC phase in QCD?
One way is to observe it through the cold compact star, which may be a hybrid of nuclear matter and the CSC matter.

The structure of such a hybrid star is implied by the phase diagram of QCD given by our model.
It is composed of a stiff nuclear outer shell and a soft CSC core.
As for the stiffness, we can estimate it by using the EoS of each part.
Our purpose is to investigate such a hybrid star to see the CSC phase in QCD. 
The possibility of such stars is examined by 
{solving the Tolman-Oppenheimer-Volkoff (TOV) equations for the compact star.
In solving TOV, we use the EoS of the CSC states which is obtained by our holographic model.} 
In this article, we concentrate our investigation on the low temperature phase of QCD.
Some discussions on the holographic modeling of the neutron star can be found in Refs. \cite{Kovensky:2021wzu, Pinkanjanarod:2020mgi, Demircik:2020jkc, Jokela:2020piw, Ecker:2019xrw, Jokela:2018ers}.

\vspace{0.3cm}
In the next section, our holographic model is given, and the EoSs of two low-temperature phases, the baryonic and CSC states, are shown.
In Sec.\,\ref{sec:TOVeq}, the TOV equations for the compact star are solved and the stability of the hybrid star is examined by
investigating the relations of the mass and the radius. 
In the last section, summary and discussions are given.

\section{A Model of Holographic QCD}

Our holographic model is constructed in the $d+1$-dimensional bulk as,
\begin{align}
   \mathcal{L} &= \mathcal{L}_\mathrm{Gravity} +\mathcal{L}_\mathrm{V} \label{action}\,, \\
   \mathcal{L}_\mathrm{Gravity} &= {\cal R} + {\frac{d(d-1)}{L^2}}\,. \label{bulk-L} 
\end{align}
The gravitational theory 
is supposed to be dual to the $d$-dimensional strongly interacting Yang-Mills (YM) theory with large $N_c$.
The typical scale of the compactified space of the original higher-dimensional gravitational theory is denoted by $L$.
The dual to the flavor part, the quark system of QCD, is denoted by $\mathcal{L}_V$. This term is given by an appropriate form for investigating the ground state of the dual QCD. 

Hereafter we consider the case of $d=5$ and concentrate on the background given at low temperature phase.
It is known as the AdS soliton solution \cite{Witten:1998zw,Horowitz:1998ha} and being written as
\begin{align}
    ds^2 &= r^2 (\eta_{\mu\nu}dx^{\mu}dx^{\nu}+f(r)dw^2) + {\frac{dr^2}{r^2f(r)}} \, ,
    \label{Soliton}
\end{align}
where 
\begin{align}
    f(r) &= 1 - \Bigl( \frac{r_0}{r} \Bigr)^5 \, , \quad r_0 = \frac{2}{5R_w} \, ,
    \label{Soliton-2}
\end{align}
and $2\pi R_w$ denotes the compactified length of $w$.
Due to this compactification of $w$, we can say that we are considering an effective 4D QCD.

\vspace{.3cm}
\subsection{Nuclear matter phase }

We consider the baryon condensed phase. 
This phase has been studied in Refs.\,\cite{Ghoroku:2012am,Ghoroku:2021fos} using the following form of $\mathcal{L}_\mathrm{V}$;
\begin{align}
    \mathcal{L}_\mathrm{V} &= - {\frac{1}{4}} F^2 - {\frac{1}{4}} {\rm tr}F^2_{SU(N_f)} \label{vector-Nf}\,, \\
    F_{\mu\nu} &= \partial_\mu A_\nu-\partial_\nu A_\mu\,,\quad
\end{align}
where $F^2=F_{\mu\nu}F^{\mu\nu}$, and $F^2_{SU(N_f)}$ denotes the two-form $SU(N_f)$ squared gauge fields.
In this case, we find an instanton configuration which is identified with the baryon.
In Ref.\,\cite{Ghoroku:2021fos}, the EoS of the dilute gas of this instanton is given. 
The way to obtain it is very briefly reviewed.
In the present case, we must add the Chern-Simon (CS) term to (\ref{action}) and modify the action such as
\begin{align}
  S &= \int d^{6} \xi \sqrt{-g} \, \mathcal{L} + S_{\rm CS} \,. 
\end{align}
This makes the $U(1)$ gauge field couple to the instanton that has a baryon-number charge. 
\footnote{Details of the CS term are shown in Ref.\,\cite{Ghoroku:2021fos}.}
Then, the free energy density $ {\cal E}$ of the instanton gas is given by using the solutions of the equations of motion of the matter system as 
\begin{align}
 S_{\rm matter} &= \int d^{6} \xi \sqrt{-g}\, \left( - {\frac{1}{4}} F^2 - {\frac{1}{4}} {\rm tr}F^2_{SU(2)}\,\right)+S_{\rm CS} \, \\
                &= -\int d^5\xi ~{\cal E}(\rho,\mu) \, ,   
\end{align}
where $\mu = A_0(\infty)$, and $\rho$ represents the instanton size, which is determined by minimizing $ {\cal E}(\rho,\mu)$. 
This procedure to find the minimum has been done numerically for each $\mu$ \cite{Ghoroku:2021fos}.
Then, the value of the free energy is determined as a function of $\mu$ as ${\cal E}(\mu)$.

Using the above numerical results for ${\cal E}(\mu)$, we find an approximate formula as a function of $\mu$;
\begin{align}
  p &= -{\cal E} = a \mu (\mu - \mu_c)\,,
  \label{pressure-1}
\end{align}
where $a=0.13$ and $\mu_c=0.17$ (see Fig.~\ref{Phase-diagram-2-1}).
Then, the energy density is given at zero temperature, $T=0$, as 
\begin{align}
   {\epsilon} = \mu{\frac{\partial p}{\partial\mu}} - p
              = a \mu^2 \, .
\label{energy}
\end{align}
As a result, we arrive at the EoS of the nuclear matter given as the instanton gas. It is
written as
\begin{align}
    p = \epsilon - \sqrt{a\epsilon} \, \mu_c \, .
\label{eos}
\end{align}
Using this EoS, the solutions of TOV equations are given in Ref.\,\cite{Ghoroku:2021fos}.
An important point to have a large sized and heavy star may be the stiffness of the constituent of the star. 
The stiffness becomes large with increasing speed of sound, $C_s$.

For the baryon phase, the speed of sound is obtained as
\begin{align}
    C_s^2 &= \frac{\partial p}{\partial \epsilon}
           = \frac{\partial p / \partial \mu}{\partial \epsilon / \partial \mu}  
           = 1 - \frac{\mu_c}{2\mu}\, .
\label{eq:cs2}
\end{align}
This formula implies the constraint on $C_s^2$, $1/2<C_s^2<1$ for $\mu_c<\mu<\infty$. 
This means that the baryonic matter is stiff compared to the quark matter in the deconfinement phase, where we will find $C_s^2\approx 1/3$ at large $\mu$ \cite{Karch:2007br}.
This point is important to construct a compact star with a core of different phases. 

\subsection{CSC phase}

We consider the CSC phase of QCD, whose holographic dual theory is given by the following action \cite{Gubser:2008px,Hartnoll:2008vx};
\begin{align}
S &= \int d^{d+1} x \sqrt{-g}\,\mathcal{L}
   = \int d^{d+1} x \sqrt{-g}\,(\mathcal{L}_\mathrm{Gravity}+\mathcal{L}_\mathrm{CSC})  \, , 
\label{bottom-up}
\end{align}
where
\begin{align}
    \mathcal{L}_\mathrm{CSC} &= - {\frac{1}{4}} F^2 - |D_{_\mu} \psi|^2 - m^2 |\psi|^2\, ,
\label{probe-L}
\end{align}
with
\begin{align}
    F_{\mu\nu} &= \partial_\mu A_\nu-\partial_\nu A_\mu \, , \quad
    D_{\mu} \psi = (\partial_{\mu}-iqA_{\mu})\psi \, .
\end{align}
This is obtained by replacing $\mathcal{L}_{V}$ in (\ref{vector-Nf}) with $\mathcal{L}_\mathrm{CSC}$, that is, the flavor part $SU(2)$ in $\mathcal{L}_{V}$ is replaced by the charged scalar part. 
The charge of the scalar, which is supposed as a diquark state, is set as $q$, which is taken as $2/N_c(=2/3)$ \cite{BitaghsirFadafan:2018iqr}.
Then, the conformal dimension of the scalar is consistent with $m^2=-4$.

The bulk configuration is given by the AdS-Soliton solution obtained from $\mathcal{L}_\mathrm{Gravity}$, and the back reactions from $\mathcal{L}_\mathrm{CSC}$ are neglected since this part is treated as a probe.
Then, the equations of motion of $A_{\mu}$ and $\psi$ are obtained by assuming that $A=A_{\mu}dx^{\mu}=\phi(r)\,dt$ and $\psi=\psi(r)$:
\begin{equation}
\psi'' + \left(\frac{6}{r}+\frac{f'}{f}\right)\psi' + \frac{1}{r^2f} \left(\frac{q^2\phi^2}{r^2}-{m^2}\right) \psi=0\;,
\label{eq1}
\end{equation}
\begin{equation}
\phi'' + \left(\frac{4}{r}+\frac{f'}{f}\right)\phi' - \frac{2q^2\psi^2}{r^2 f}\phi=0\;.
\label{eq2}
\end{equation}
Since $f(r)$ vanishes at $r=r_0$, Eqs.\,(\ref{eq1}) and (\ref{eq2}) should be solved under the following conditions;
\begin{align}
    \phi'(r_0) = \frac{2q^2\psi^2(r_0)}{5r_0}\phi(r_0)\, , \quad 
    \psi'(r_0) = -\frac{1}{5r_0} \left(\frac{q^2\phi^2(r_0)}{r_0^2} - m^2\right) \psi(r_0)\,, 
\label{bc1}
\end{align}
to evade the singularity.
Here, we notice that the boundary condition (\ref{bc1}) allows the solution of $\phi(r_0)\neq 0$.
The details of the solutions of the above equations can be seen in Ref.\,\cite{Ghoroku:2019trx}, so we omit them here.

\vspace{.3cm}
\subsubsection{On-shell Euclidean action and EoS}

We estimate the free energy, which can be obtained by the on-shell action. 
The Euclidean action is separated to the bulk and probe parts as
\begin{align}
 S^{\rm E}  &=  -\int d^{d+1} x \sqrt{-g}\,\mathcal{L} = S_{\rm bulk}^{\rm E} +S_{\rm probe}^{\rm E}\,, 
\end{align}
where
\begin{align}
    S_{\rm bulk}^{\rm E} &= -\int d^{d+1} x \sqrt{-g}\,\mathcal{L}_\mathrm{Gravity} = -r_0^5 {\frac{4\pi}{5r_0}}{\frac{1}{T}} V_3 \, , \\
    S_{\rm probe}^{\rm E} &= -\int d^{d+1} x \sqrt{-g}\,\mathcal{L}_\mathrm{CSC} \, ,  
\end{align}
with $V_3=\int d^{d-3} x$. 


The probe action $S_{\rm probe}^{\rm E}$ is separated 
into two parts;
\begin{align}
 {\frac{S_{\rm probe}^{\rm E}}{V_5}} = \hat{S}_{\psi}^{\rm E} + \hat{S}_{\phi}^{\rm E}\,, 
 \label{eq:action_prob}
\end{align}
where $V_5(=\int d^5\xi)$ denotes the volume of the boundary space-time.
Then their on-shell parts are estimated by using the solutions, $\psi$ and $\phi$, of the Eqs. (\ref{eq1}) and (\ref{eq2}).
As for the first term, the $\psi$-dependent part, we see
\begin{align}
   \hat{S}_{\psi}^{\rm E}
       &= -\int dr\sqrt{-g}  \left( - |D_{_\mu} \psi|^2 - m^2 |\psi|^2 \right)  \\
       &= \int dr \sqrt{-g}  \left(g^{rr}{\psi'}^2+q^2 A_0^2\psi^2g^{00} + m^2\psi^2\right) \\
       &=  \int dr \sqrt{-g} \left[ -\frac{1}{\sqrt{-g}} \partial_r \left(\sqrt{-g}(g^{rr}{\psi'}\right) + q^2 A_0^2 \psi^2 g^{00} + m^2\psi\right]\psi\,
           \nonumber \\
       & \hspace{2.3cm} + \left[ \sqrt{-g} g^{rr} \psi' \psi \right]_{r_0}^{\infty} . 
\label{eq-3}
\end{align}

The integrand part in Eq.\,(\ref{eq-3}) vanishes due to the equation of motion (\ref{eq1}). By the boundary term, Eq.\,(\ref{eq-3}) is then estimated as
\begin{align}
       \hat{S}_{\psi}^{\rm E} &= \left[ \sqrt{-g} g^{rr} \psi' \psi \right]_{r_0}^{\infty}
                               = \left[ r^6f(r)\psi\psi' \right]_{r_0}^{\infty} \label{eq-5} \
                               = 0\,  ,
\end{align}
where we used 
\begin{align}
  f(r_0) = 0 \, , \quad \left. \psi(r)\right|_{r\to \infty} ={\frac{C}{r^4}} + \cdots . 
\end{align}

For the second term at the r. h. s. of Eq.\,(\ref{eq:action_prob}), we see
\begin{align}
       \hat{S}_{\phi}^{\rm E} &= -\int dr\sqrt{-g} \left( - {\frac{1}{4}} F^2 \right)\,  \\
                              &= -\int dr \sqrt{-g} \left(-{\frac{1}{2}}g^{00}g^{rr}{\phi'}^2\right)\,  \\
                              &= \int_{r_0}^{\infty} dr \left( q^2r^2\psi^2A_0^2 \right) -{\frac{3}{2}}\bar{d}\mu\,, 
\end{align}
where we assumed
\begin{align}
    \left. A_0(r)\right|_{r\to\infty} =\mu - {\frac{\bar{d}}{r^3}} + \cdots\,.
\end{align}
Then, the pressure of the diquark superconducting gas is given as
\begin{align}
  p_s = \frac{3}{2} \bar{d} \mu - \int_{r_0}^{\infty} dr \left( q^2r^2\psi^2A_0^2 \right)\,.
\end{align}
We can see that the pressure given above is positive by rewriting it in a simple form as
\begin{align}
       p_s = \int_{r_0}^{\infty} dr \left( {\frac{1}{2}} r^4f {\phi'}^2 \right) \,.
\label{p-simple}
\end{align}

Here $p_s$ represents the pressure of the matter part.
On the other hand, that of the gravity part is omitted since the gravitational background is common to the two different matter systems considered here due to the probe approximation.
Then the phase transition can be examined by comparing the free energies of the different matter parts coupled to the common gravitational background.

\subsubsection{Numerical result and phase transition}

\begin{table}[htbp]
\caption{The pressure $p = p(\mu)$ versus $\mu$ for the low temperature CSC phase.
The numerical estimations are given for $r_0=1$.} 
\begin{center}
\begin{tabular}{ll}
\multicolumn{1}{c}{$\mu$} & \multicolumn{1}{c}{$p$} \\
\hline 
4.678 & ~0.03465 \\
4.95202 & ~0.4825 \\
5.36667 & ~2.2515 \\
6.07538 & ~8.90935 \\
6.82273 & 22.3817 \\
7.63461 & 46.8952 \\
8.55023 & 92.0457 \\
\hline
\end{tabular}\label{table-1}
\end{center}
\end{table}

\begin{figure}[htbp]
\vspace{.3cm}
\begin{center}
\includegraphics[width=10.0cm]{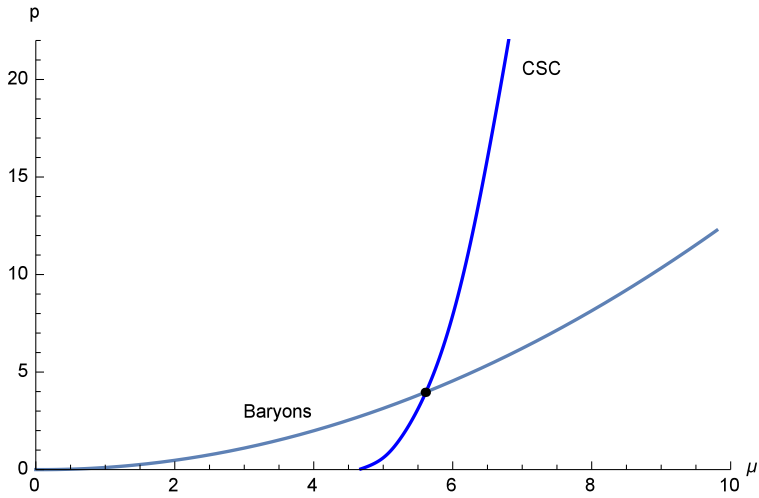}
\caption{  The $\mu$-$p$ curves for the phases of CSC and baryon phases.
%
%
\label{Phase-diagram-2-1}}
\end{center}
\end{figure}

The numerical estimations for $p_s=p(\mu)$ at several points of $\mu$ are shown in Table \ref{table-1}. 
The resultant $p(\mu)$ is plotted in Fig. \ref{Phase-diagram-2-1} and compared with $p(\mu)$ obtained for the baryon phase.
We find the phase transition point between these two phases.
The critical point is shown by the dot in Fig. \ref{Phase-diagram-2-1}.
For the baryon phase, the curve represents $p=0.13 \mu(\mu-0.17)$ as given by Eq.\,(\ref{pressure-1}).
For the CSC phase, the curve shows a smoothly connected one of the points given in Table \ref{table-1}. 
{As for the numerical values of $\mu$ and $p$ appeared in the above Table 1 and Fig. 1, 2 and 3 given below, they can be rewritten 
by dimensionful values as, $p e_0$ and $\mu e_0^{1/4}$
where $e_0=8.52\times 10^4$ MeV/fm$^3$ and  $e_0^{1/4}=0.896$ GeV. The parameter $\epsilon_0$ is
introduced to solve the TOV equations \cite{Ghoroku:2021fos}. Here we do it when we need the dimensionful values. }


\begin{figure}[htbp]
\vspace{.3cm}
\begin{center}
\includegraphics[width=12.0cm]{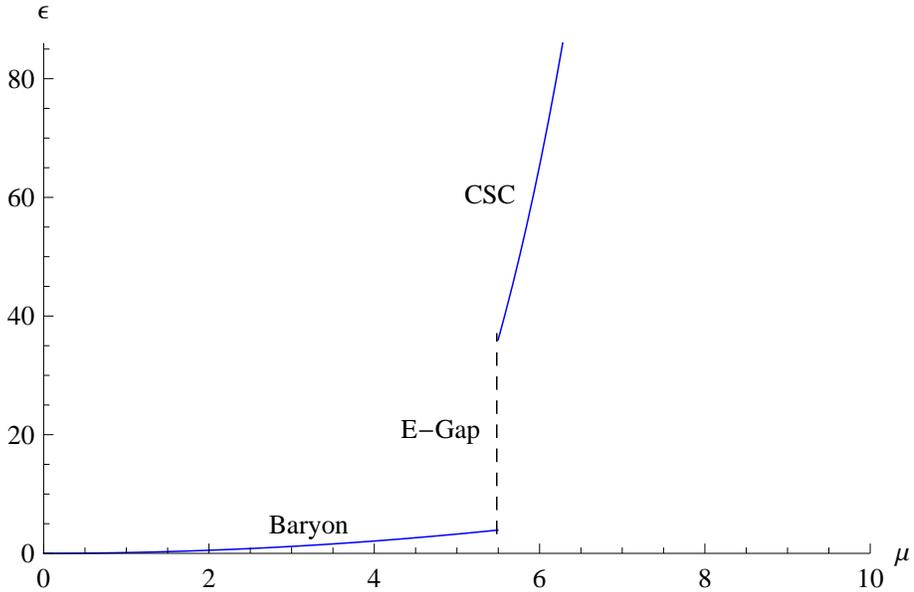}
\caption{  The $\mu$-$\epsilon$ curves for the phases of CSC and baryon phases.
\label{Phase-diagram-2-2}}
\end{center}
\end{figure}

\vspace{.3cm}
When the $p(\mu)$ in the CSC phase is replaced by an analytical function of $\mu$, then we can proceed with the analysis more smoothly. 
Here, we assume that such a $p(\mu)$ is written in a power series form, 
\begin{align}
    p_n(\mu) &= \sum_{i=0}^n a_i \, \mu^i\,.
\end{align}
The value of $n$ is related to the space-time dimension of the theory on the boundary.
Then, $n=4$ is expected here, this is seen from the analysis given in the type IIB model \cite{Karch:2007br}, where an analytic form of free energy is obtained at $T=0$. 
In our present model, the solution for $n=4$ is obtained as
\begin{align}
    p_4 &= a_0+a_1\mu+a_2\mu^2+a_3\mu^3+a_4\mu^4\,,
\end{align}
where
\begin{align}
    &a_0=48.2717\,, ~~a_1=-24.7298\,, ~~a_2=5.18881\,,~~a_3=-0.754903\,, ~~a_4=0.0650574\,. \nonumber
\end{align}

Then, we can show by using this $p(\mu)$ for $n=4$ that the above transition is first-order since the energy density $\epsilon$ has a gap at the transition point, $(\mu,\,p)=(5.60593, \,3.96155)$ as shown in Fig. \ref{Phase-diagram-2-2}. The curves for $\epsilon$ are obtained by using the formula 
\begin{align}
\epsilon = \mu {\frac{\partial p}{\partial\mu}} -p. \label{energy1}
\end{align}

The dotted line shows the discrete energy gap between the values on the two phases at {the critical $\mu$, $\mu_\mathrm{t} = 5.60593$}. 
This implies the first-order phase transition between (c) and (a-B) phases.
{We give a comment here on the energy density of the baryon matter near this critical point. 
It is given as $\epsilon(=0.13\times 5.60593^2\times \epsilon_0)=348$ GeV/fm$^3$ at $\mu=\mu_\mathrm{t}$.
This implies that the density of the baryon
(or the instanton) is very large and we should modify our model of dilute gas approximation
in this region. One way to avoid this difficulty is to reset the value of $\epsilon_0$ to a smaller value.
In this case, however, the size of the neutron star $R$ is extended to a larger value. 
Further discussion on this point is postponed to a future problem. }

We notice that $p(\mu)$ for $n=5$ can be considered as another possible pressure form since the present holographic model has one extra compactified dimension on the boundary.
By giving an appropriate form of $p_5$, it will be possible to study the availability of our model by comparing various quantities obtained in terms of $p_5$ with the ones given by $p_4$.
However, this investigation is remained as the future work.

\begin{figure}[htbp]
\vspace{.3cm}
\begin{center}
\includegraphics[width=11.0cm]{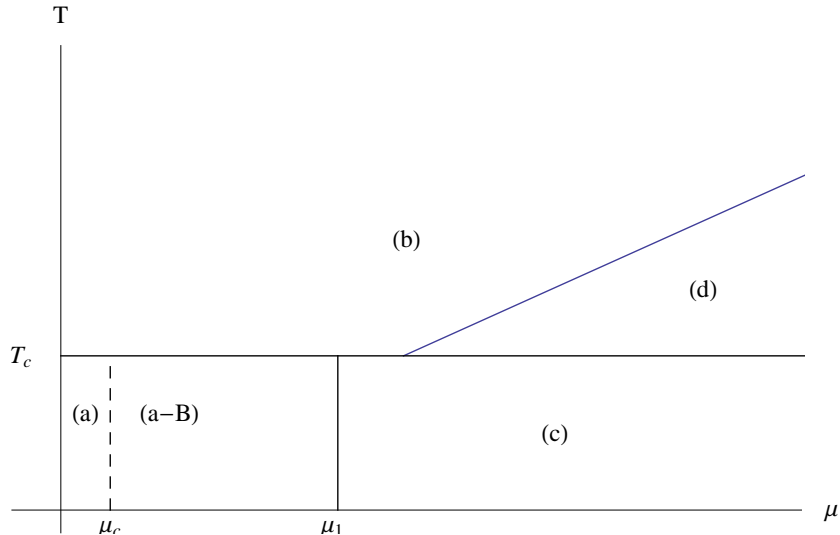}
\caption{ Schematic figure of the phase diagram for AdS-Schwarzschild ($T>T_c\simeq 0.4$) and AdS soliton confinement background ($T<T_c$), with $q=2/3$, $m^2=-4$.
{Using present mass unit $\epsilon_0^{1/4}=0.899$ GeV \cite{Ghoroku:2021fos}, we have $T_c=0.36$ GeV, $~\mu_c=0.152$ GeV and $\mu_t=5.04$ GeV. }
\label{Phase-diagram-1}}
\end{center}
\end{figure}

\subsection{Phase diagram}

In our model, the phase diagram is given in Fig.~\ref{Phase-diagram-1} in the $\mu$-$T$ plane. 
The horizontal line, $T={\frac{5}{4\pi}}=T_c$, shows the critical line between the confinement $(T<T_c)$ and deconfinement $(T>T_c)$ phases. 
The critical point, $T_c$, is obtained by using Eq.\,(\ref{bulk-L}).
As explained above, the other transition lines are given by adding $\mathcal{L}_V$ as a probe in the two-form of
background configurations given by $\mathcal{L}_\mathrm{Gravity}$.
Then we find two CSC phases (c) and (d). While
it is an interesting point how they are different from each other, we here concentrate on the low temperature side ($T<T_c$) of 
baryon condensed phase (a-B) and the CSC phase (c).

This phase diagram suggests us a cold compact star with the core of CSC matter. Noticing that the pressure of the matter 
monotonically increases with the chemical potential $\mu$. Then we have a natural composition of the star, which is made of nuclear matter as outer shell
and a CSC core inside. In this construction, the pressure decreases from the maximum center value to the vanishing surface value.  
In the next section, we investigate the possibility of this type of cold star.

\vspace{.5cm}
\section{Compact Star with CSC Core}
\label{sec:TOVeq}

\subsection{Phase Transition and TOV solutions}

In case that the pressure is given as a function of the chemical potential, the TOV equation can be rewritten as follows:\footnote{{The meanings of $m$ and $r_0$ below are different from those in the other sections.}}

\begin{align}
    \frac{d\mu}{dr} &= -\mu\frac{m + 4\pi r^3p}{r(r-2m)}\ ,\\
    \frac{dm}{dr} &= 4\pi r^2\epsilon\ .
\end{align}

Inside a star of our model, the two phases contact each other through a spherical boundary at $r=r_\mathrm{b}$, where the pressure is continuous.
However, the chemical potential keeps its equilibrium by satisfying that $\mu_\mathrm{B}=N_c\mu_\mathrm{q}$ at the boundary.
Therefore, we should interpret $\mu$ as \textit{the reduced chemical potential};
\textit{i.\,e.\,}, $\mu=N_c\mu_\mathrm{q}$ for $r\le r_\mathrm{b}$ and $\mu=\mu_\mathrm{B}$ for $r\ge r_\mathrm{b}$.
Even a case of two-phase star, TOV equation can be solved with a set of initial conditions given at a certain point in the core.
One of them is $\mu_1=\mu(r_1)$, which is the initial value of the chemical potential at a point very near the center, $r=r_1$; another is $\epsilon_1=\epsilon(\mu_1)$, which is the energy density within the radius $r_1$ and is well approximated by $m(r_1)=(4\pi r_1^3/3)\epsilon_1$.
This means that the initial value $\mu_1=\mu(r_1)$ governs both parts of the star in the following way:
First, numerical integration for the core part is terminated where $\mu$ reaches  the critical value $\mu_\mathrm{t}$, and the core radius $r_\mathrm{b}$ is immediately determined.
Second, for the outer shell, the set of initial values is given at $r=r_\mathrm{b}$ (the boundary between the two phases), and it consists of $\mu_\mathrm{t}$ and $\epsilon(\mu_\mathrm{t})$.
This time, the numerical integration is terminated where $p$ vanishes, and finally the radius of the star ($R$) is determined.

\begin{figure}[htbp]
\centering
\includegraphics[width=12cm]{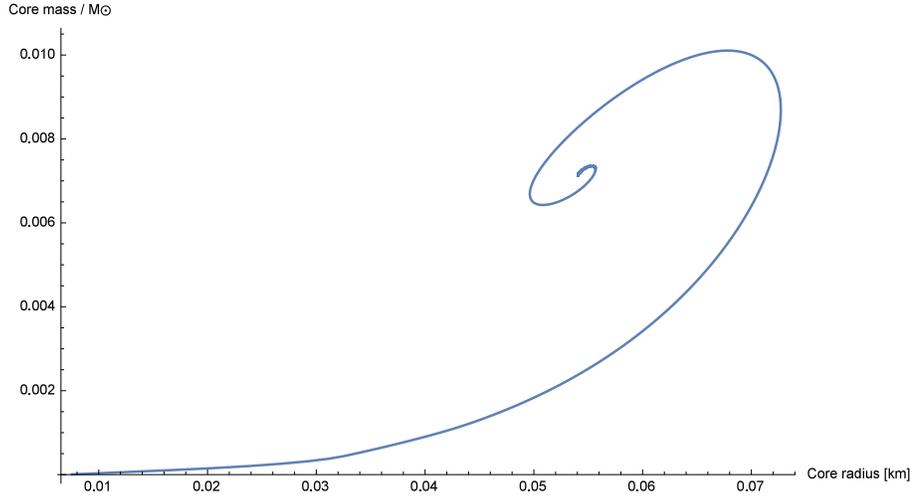}
\caption{The $M$-$R$ relation for the core part.
The curve covers the region of $\mu\ge\mu_\mathrm{t}$, and starts from the lower-left endpoint. The marks ($\mathrm{A}$ to $\mathrm{D}$) correspond to those in Fig.\,\ref{fig:outershell2c}.}
\label{fig:corepart2c}
\end{figure}

\begin{figure}[htbp]
\centering
\includegraphics[width=12cm]{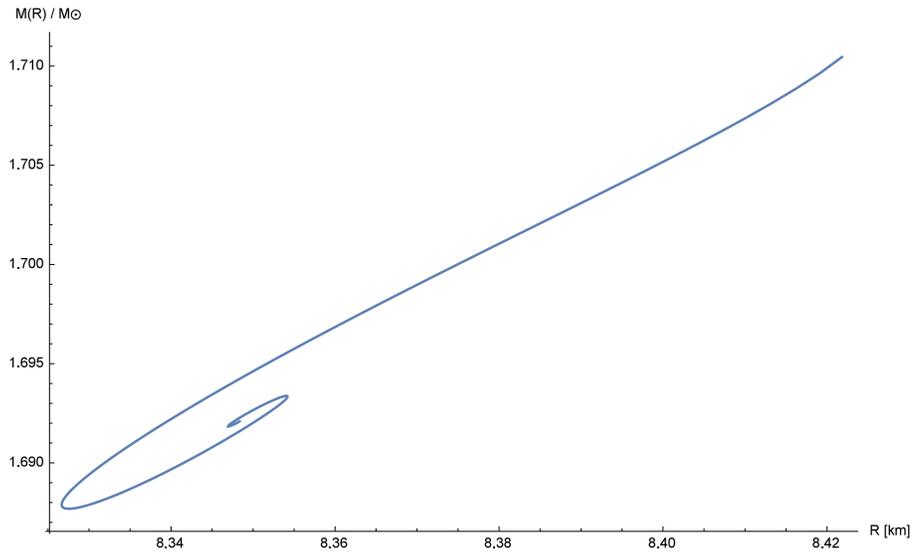}
\caption{The $M$-$R$ relation for the whole star.
The upper-right endpoint corresponds to the minimum chemical potential ($\mu=\mu_\mathrm{t}$) at the center of the core.
The curve does not extend from the endpoint, in the present two-phase model.}
\label{fig:outershell2c}
\end{figure}

There are two types of $M$-$R$ relations, for the core part and the whole star.
In each relation, the physical dimensions are recovered by introducing a typical length $r_0$ and by the transformation $(r,\,m,\,p,\,\epsilon)\rightarrow(r_0r,\,m_0m,\,p_0p,\,\epsilon_0\epsilon)$ with
\begin{equation}
m_0=\frac{r_0c^2}{G}\ ,\quad
p_0=\epsilon_0=\frac{c^4}{r_0^2G}\ .
\end{equation}
For $r_0=3.00\,\mathrm{km}$, it is calculated as $m_0=2.03M_\odot$ and
$p_0=\epsilon_0=1.34\times10^{37}\,\mathrm{J/m^3}$.
Assuming that stable stars with CSC core exist, we show both relations by Figs.~\ref{fig:corepart2c} and \ref{fig:outershell2c}.
In Fig.\,\ref{fig:corepart2c}, we show the behavior of the CSC core as a function of the core mass and its radius.
In contrast, Fig.\,\ref{fig:outershell2c} represents the $M$-$R$ relation for the whole star, but we only plot the solution of TOV equation on the figure restricting the cases with the CSC core. If the CSC core vanishes inside the compact star, the result must be that with single phase.

\begin{figure}[htbp]
\centering
\includegraphics[width=12cm]{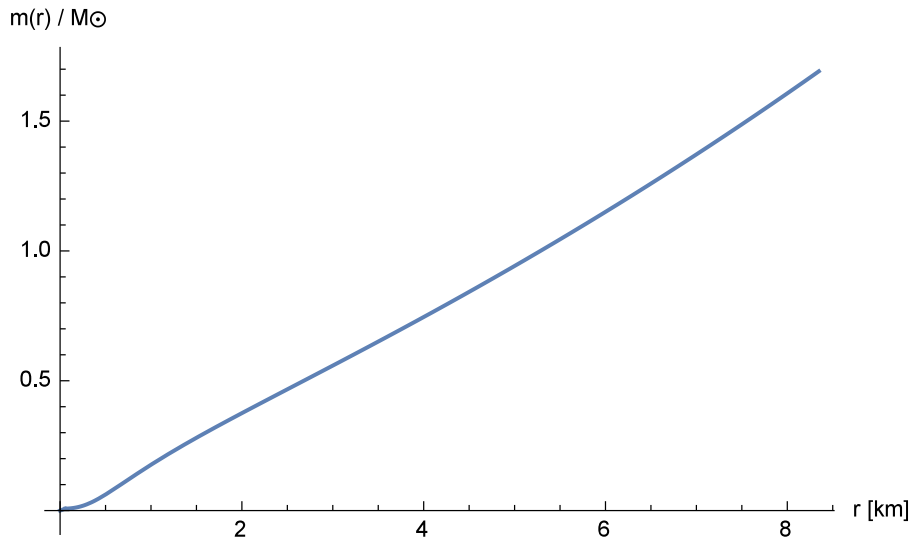}
\caption{An example of the mass inside $r$, in case of $\mu(r=0)=55.56$.
The phase changes at $r=0.055\,\mathrm{km}$, the core mass is $0.733\times10^{-2}\,M_\odot$,
the radius of star is $R=8.35\,\mathrm{km}$, and the total mass of star is $M=1.69 M_\odot$.}
\label{fig:m_inside_r3}
\end{figure}

We also give an example of mass accumulation in a two-phase star, by Fig.~\ref{fig:m_inside_r3}.
As seen from the figures, even if a two-phase star exists, the scale of the core is very small compared to that of the outer shell, and the total mass of the star is contributed almost by the outer shell.
However, unlike a single-phase star, the chemical potential in the outer shell (nuclear part) is restricted below $\mu_\mathrm{t}$, which might strongly control the upper limit of $M$.
Beyond the upper-right endpoint of Fig~\ref{fig:outershell2c}, the $M$-$R$ curve extends to a part of a single-phase curve as calculated
in Ref.~\cite{Ghoroku:2021fos}.
Conversely, the $M$-$R$ curve of the single-phase star changes the branch at $\mu=\mu_\mathrm{t}$ to that of two-phase star.

\subsection{Stability of the Hybrid Star}
The next point to be examined is the stability of the two-phase star, the hybrid of nuclear matter and CSC matter.
The stability of such a star as 
the solutions of the TOV equations can be read from the $M$-$R$ relation shown in the Fig. \ref{fig:outershell2c}. 
As for the stability of these solutions, it is possible to check the following points of the $M$-$R$ relation.
The first point is that the solution has an unstable mode when it is in a region of the curve of $M$-$R$ relation where the state of the star satisfies the condition
\cite{BitaghsirFadafan:2020otb}  
\beq
\frac{\partial M}{\partial \epsilon_c}<0\, ,
\eeq
where $\epsilon_c$ denotes the central energy density, $\epsilon_c=\epsilon |_{r=0}$. 
We notice here that $\epsilon$ increases monotonically with $\mu$.

Second, we check the $M$-$R$ curve at the extremum points by the Bardeen, Thorne and Meltzer (BTM) criteria \cite{Alford:2017vca,bardeen1966catalogue}
{ which are given by the following statements. (i) At each extremum where the $M$-$R$ curve rotates counter-clockwise with increasing $\epsilon_c$, one unstable mode appears. (ii)  At each extremum where the $M$-$R$ curve rotates clockwise with increasing $\epsilon_c$, one unstable mode becomes stable.  
}

{In Fig. \ref{fig:outershell2c}, the point of the solution moves on the curve from the upper endpoint to the point $\mathrm{B}$ and $\mathrm{C}$ with increasing $\epsilon_c$. Then the solutions up to $\mathrm{D}'$ on the upper curve are unstable.
In the next, other unstable modes appear for the solutions across the extrema $\mathrm{B}'$ and $\mathrm{C}'$. Any solution on the curve of Fig. \ref{fig:outershell2c} is therefore unstable.
Then, we cannot find any stable solution of a star with CSC core. 
This result implies that it seems to be very difficult to observe the CSC state in the star.
}


\begin{figure}[htbp]
\centering
\includegraphics[width=12cm]{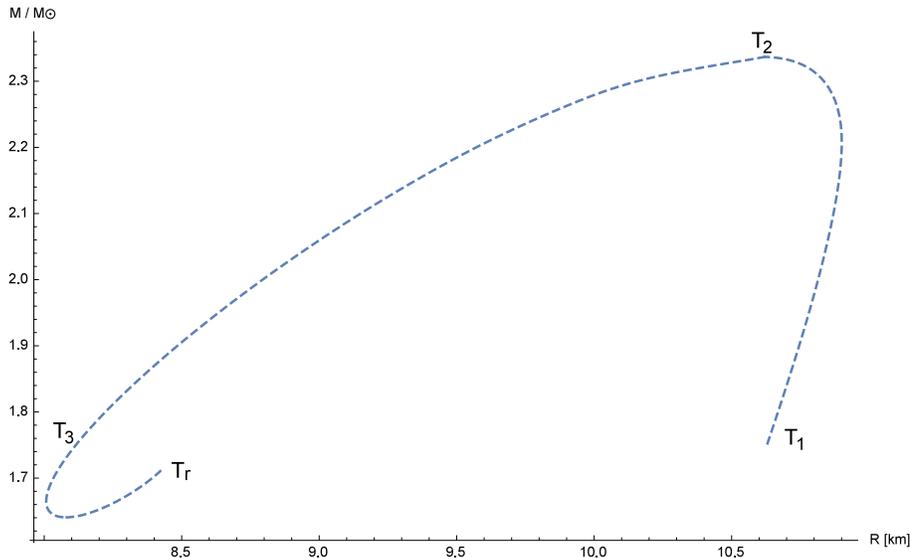}
\caption{{The $M$-$R$ relation for the nuclear matter star. Notice that the small center $\mu$ part of the curve has been 
given in Ref.~\cite{Ghoroku:2021fos}.}
\label{baryon1x}}
\end{figure}



\begin{figure}[htbp]
\centering
\includegraphics[width=12cm]{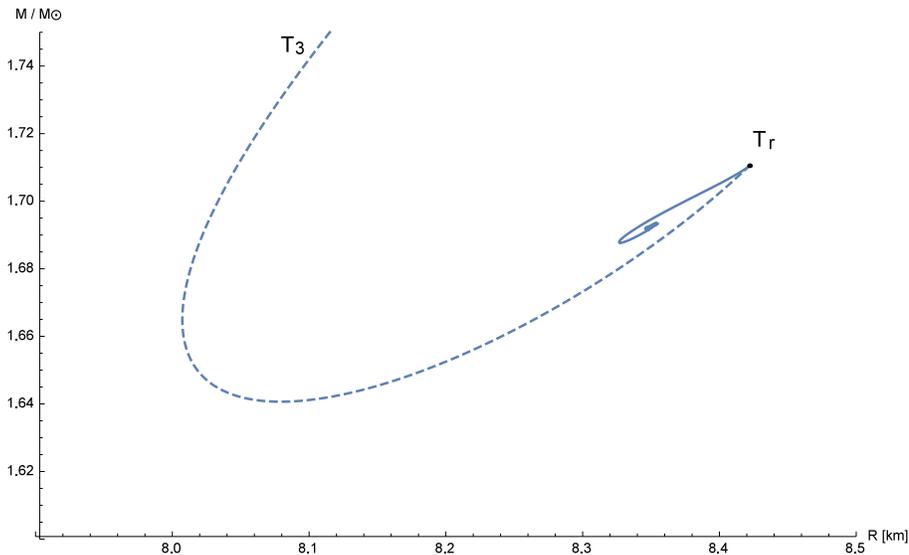}
\caption{{The dotted (solid) curve repressents $M$-$R$ relation for the nuclear matter star (hybrid star). 
The two $M$-$R$ curves are connected at the point $\mathrm{T_r}$. }
\label{baryon2x}}
\end{figure}

{In order to understand the situation 
of the hybrid star and the star of pure nuclear matter, we combine the $M$-$R$ relation of 
the hybrid star shown in the Fig. \ref{fig:outershell2c} to the one of the nuclear star. The latter is given in the Fig. \ref{baryon1x}, where 
the value of the central chemical potential, $\mu(r=0)$, 
increases from the point $\mathrm{T}_1$ to $\mathrm{T_r}$ along the curve. In the present model,
the curve ends at $\mathrm{T_r}$ since the CSC core will be generated at this point.
In other words, the point $\mathrm{T_r}$ 
denotes the transition point to the hybrid star where the center chemical potential arrives at the critical point, 
the upper end point, $\mu(r=0)=\mu_\mathrm{t}$, of the hybrid star curve.
In the case of the nuclear star,
the stable solutions exist in the region from $\mathrm{T}_1$ to $\mathrm{T}_2$, and the one between $\mathrm{T}_2$ and $\mathrm{T_r}$ are unstable as we can understand from
the rule to find the unstable radial fluctuation mode mentioned above.
Then the two curves in the Fig. \ref{fig:outershell2c} and Fig. \ref{baryon1x} 
are connected at $\mathrm{T_r}$ as shown in the Fig. \ref{baryon2x}.
}

The reason why it is difficult to find a stable CSC core is that the outer nuclear shell covers 
too large chemical potential part up to $\mu_\mathrm{t}\sim 5.6$. 
The stable solution for pure nuclear matter is however obtained in the region $\mu<0.4$ \cite{Ghoroku:2021fos}. 
Then to find a stable star with the CSC core we must consider a model which gives a small value of $\mu_\mathrm{t}$.
The resolution of this point is an open problem here.

\vspace{.5cm}
\section{Summary and Discussions}

Based on a bottom up holographic model of QCD, a hybrid star of the nuclear matter and the CSC matter is studied.
The analysis is executed by a probe approximation in the low temperature confinement phase. Then the back reactions from the flavor matter sector 
to the color gauge part are neglected. 
In this case,  when $\mu$ increases from zero, we find a baryon condensed phase when $\mu$ reaches the critical point $\mu_\mathrm{c} \sim 0.17$. 
Then, at the second critical point, $\mu_\mathrm{t} \sim 5.6$, this baryonic matter changes to the CSC phase 
via the first-order phase transition. The EoSs of these connected two phases 
are given holographically, and we find that, in both phases, the pressure increases monotonically with $\mu$. 
Then it seems to be natural to suppose a star which is composed of CSC core and outer nuclear shell. 

In order to see the possibility of such a star, the relations of the mass and the radius of this hybrid star is studied. 
The mass-radius curve is obtained by solving the TOV equations.
The resulting curve provides us with information on the existence of unstable modes of radial fluctuations for the solutions of the TOV equation.
However, we cannot find any stable hybrid solution on the curve studied here. This implies that it would be 
{impossible} to find the CSC matter in the compact cold star.

\vspace{.5cm}

We give here several comments on a trial to find the CSC core in a cold star from a slightly different viewpoint as follows:

(I) {In the present study, the TOV equations are solved with two phases.
The existence of the CSC core in the figure of the $M$-$R$ relation is restricted to a very tiny region, and thus there is no significant effect
  even if we consider the crossover scenario; see Appendix for some discussions on the crossover scenario 
  and model properties.
}

 (II) 
In this paper, we neglected the interaction effects among baryons. Once we take into account such effects, the baryons start to overlap more easily and the transition to (color-superconducting) quark matter might occur at smaller chemical potential. Then the effect of the CSC core for the $M$-$R$ curve could be seen.
This point remains an open problem at present. 

(III) We should consider other types of holographic models which give smaller $\mu_\mathrm{t}$.
One possible way is to use the model which can represent deconfined state as a background.
Since we only consider the confined state in the present study as a first attempt to investigate the two-phase compact star, the CSC core tends to shrink.
Actually, the Reissner-Nordstrom (RN) charged black hole solution can provide a smaller $\mu_\mathrm{t}$~\cite{Basu:2011yg,Ghoroku:2019trx,Nam:2021qwv} and thus this possibility may be feasible.
We will discuss this elsewhere. 

(IV) {Finally, related to the RN background for the bulk configuratioon, we comment on the backreaction. }
{If the backreaction is taken into account, as is shown in Ref. [12],
the AdS Schwarzschild background, which corresponds to the deconfinement phase, is replaced to the RN background. As the result, the critical chemical potential in the confinement-deconfinement phase transition becomes smaller and we expect the earlier onset of the quark matter inside the neutron star. In this case, however, quark matter is not color-superconducting, rather exists as the deconfined free one.} {
One way to keep CSC phase, in this case, may be to take into account of the higher power terms of the curvatures \cite{Nam:2021qwv}.}

\section*{Acknowledgments}
M. T. would like to thank for fruitful discussions during the APCTP focus program ``QCD and gauge/gravity duality''.
We also would like to appreciate Y. Izumi for useful discussions.

\newpage
\vspace{.5cm}

\appendix
\noindent{\bf\Large Appendix}

\section{Smooth transition scenario}

In the main text, the TOV equations with two phases were studied to obtain the mass-radius relation of compact star.
As seen from Fig. 2, it is natural to consider the first-order phase transition between nuclear matter and color superconducting 
quark matter phases.
On the other hand, since our study here is limited to the probe approximation, we cannot precisely tell about the genuine phase
transition patterns. Therefore, it is possible to consider the case where those two phases are smoothly connected.
Such a case is studied here and we call the case the smooth phase transition scenario.
In this section, we discuss the crossover scenario and its effects on the speed of sounds and the trace anomaly; for example, see Ref.\,\cite{Baym:2017whm} and references therein for the details of the crossover scenario.

\subsection{Speed of sound and trace anomaly}

In the smooth phase-transition scenario, let us define the pressure as
\begin{equation}
p=(1-f)p_\mathrm{NM}+fp_\mathrm{CSC},
\label{crossover eos1}
\end{equation}
where $p_\mathrm{NM}$ ($p_\mathrm{CSC}$) is the pressure for the nuclear matter (the quark matter with color superconducting), and $f$ is the interpolation function given as follows:
\begin{equation}
f=\frac{1}{2}\left [ 1+\tanh\left (\frac{\mu-\mu_*}{N_s} \right ) \right ].
\label{crossover eos2}
\end{equation}
Here $\mu_*$ shows the cross-point of the two curves for the pressures, $p_{NM}$ and $p_{CSC}$.
$N_s$ is a constant that controls how to interpolate two phases. The limit $N_s \rightarrow 0$ corresponds to the
step function, i.e.,
\begin{equation}
\frac{1}{2}\left [ 1+\tanh\left (\frac{\mu-\mu_*}{N_s} \right ) \right ] \quad \longrightarrow \quad \Theta(\mu-\mu_*),
\label{crossover eos3}
\end{equation}
where $\Theta(x)=0$ for $x<0$ and $\Theta(x)=1$ for $x>0$.
If we replace the function by the step function, the model is corresponding to the bottom-up model used in the main text.

\begin{figure}[htbp]
\vspace{.3cm}
\begin{center}
\includegraphics[width=10.0cm]{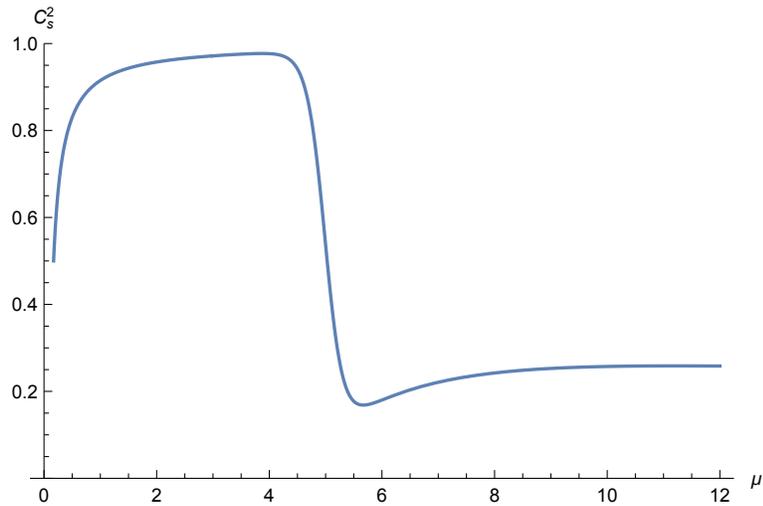} 
\caption{Speed of sound in the smooth phase transition scenario.}
\label{fig:cs2}
\end{center}
\end{figure}

\begin{figure}[htbp]
\vspace{.3cm}
\begin{center}
\includegraphics[width=10.0cm]{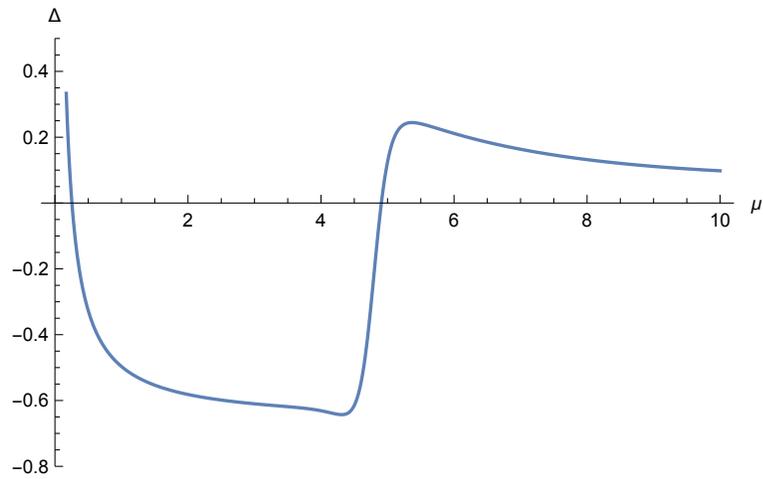}
\caption{Trace anomaly in the smooth phase transition scenario.}
\label{fig:trace_anomaly}
\end{center}
\end{figure}

Using Eq.\,(\ref{crossover eos1}), one can evaluate the speed of sound $C_s^2$ as the function of chemical potential, which is shown in Fig. \ref{fig:cs2}.
Since our nuclear matter EOS is relatively stiff, the value of $C_s^2$ in the nuclear matter phase quickly approaches 1. 
Then in the crossover region where nuclear matter
is smoothly connected to color-superconducting quark matter, $C_s^2$ drops down.
Finally, in the color superconducting phase, the value of $C_s^2$ is asymptotically close to 1/4. We can also compute the trace anomaly defined as
\begin{equation}
\Delta = \frac{1}{3}-\frac{p}{\epsilon},
\label{eq:trace_anomaly_measure}
\end{equation}
which measures the conformality of the system. 
This is the convenient measure of the trace anomaly used in Refs.~\cite{Gavai:2004se,Marczenko:2022jhl,Fujimoto:2022ohj}.
Figure \ref{fig:trace_anomaly} demonstrates the trace anomaly in the smooth phase transition scenario, where $\Delta$ ranges from $-2/3$ to $1/3$.
The asymptotic value of $\Delta$ is $1/12$ which deviates from zero.
Note here that the authors of \cite{Fujimoto:2022ohj} have discussed both the speed of sound and the trace anomaly using the EOSs extracted from the neutron star observations. It is interesting to compare their results with those obtained here; see the next subsection.

\vspace{.5cm}
\subsection{Properties of Our Model}

In the present study, we employ the $3+1+1$ dimensional bottom-up holographic model as a QCD effective model, and thus there should be a valid region where we can treat it as the $3+1$ dimensional effective model; the extra dimension should not affect the thermodynamics in the region.

One of the promising ways to estimate the validity region of the effective model is to use the $M$-$R$ relation because we have several observation data of neutron stars where EoS plays a crucial role.
Actually, we already have the following important restrictions:
\begin{itemize}
\item From Shapiro delay measurement, two solar mass ($2M_\odot$) neutron star is observed and thus $M$-$R$ curve must reach the value. See Refs.\,\cite{Demorest:2010bx,Antoniadis:2013pzd}.
\item From GW170817 via the gravitational wave observation, we have the radius constraint $9.0$ km $< R < 13.6$ km for $M = 1.4M_\odot$.
See Refs.\,\cite{Annala:2017llu,De:2018uhw,Tews:2018iwm}.
\item From the analysis of GW170817 with no detection of a relativistic optical counterpart, the upper bound of $M$ for cold spherical neutron stars is estimated as the range $2.15 M_\odot - 2.26 M_\odot$.
See Ref.\,\cite{Shibata:2017xdx}.
\end{itemize}
Unfortunately, such data are not sufficient to strictly restrict EoS, and thus we still have several unclearness's.

For the evaluation of the validity of our holographic model, the measure of the trace anomaly ($\Delta$) and the sound of speed ($C_s^2$) are additional interesting quantities.
The quantity $\Delta$, which is defined in Eq.\,(\ref{eq:trace_anomaly_measure}), is somewhat related to the absolute value of $p$ and $\epsilon$, and $C_s^2$, which is defined in Eq.\,(\ref{eq:cs2}), relates to the $\mu$-derivative of $p$ and $\epsilon$.
These indicate that we may check the validity region of our holographic model as an QCD effective model from the behavior of $\Delta$ and $C_s^2$, in principle, because both are directly affected by the behavior of EoS as a function of $\mu$. 
Since we know the tendency of the asymptotic limit of $\Delta$, such as $\Delta \to 0$ in QCD, we can use it as an indicator that our model can treat as the $3+1$ dimensional effective model; for example, see Ref\,.\cite{Fujimoto:2022ohj} and references therein for discussions of $\Delta$ in QCD.

The measure of the trace anomaly $\Delta$ tends to $0$ in the large $\mu$ limit in the four-dimensional model, but deviates from $0$ in a model with an extra dimension because the extra dimension can contribute to the measure in the region.
This may indicate the acceptable energy regime of the extra-dimensional model as the four-dimensional model.
Figure\,\ref{fig:trace_anomaly} shows the $\mu$-dependence of $\Delta$ of our holographic model.
We can see that $\Delta$ approaches $1/12$ when $\mu$ becomes large, and thus such a large $\mu$ region is not reliable.
In the intermediate $\mu$ region, $\Delta$ becomes negative, and after it shows the peak structure.
This behavior has been discussed as a plausible scenario in Refs.\,\cite{Marczenko:2022jhl,Fujimoto:2022ohj}.
Of course, the intermediate $\mu$ region is difficult to clarify at present because we do not have an exact result of QCD. 

At present, we can say that our holographic model is not inconsistent with several restrictions from the viewpoint of the $M$-$R$ relation, the trace anomaly, and the speed of sound.
A more detailed check will be discussed elsewhere.

\newpage

\bibliographystyle{elsarticle-num}
\bibliography{ref}

\end{document}